\documentstyle[11pt,twoside,epsfig]{article}

\newcommand{\beq}{\begin{equation}}
\newcommand{\eeq}{\end{equation}}

\newlength{\dinwidth}
\newlength{\dinmargin}
\begin{document}

\begin{titlepage}
DESY 94-231\\
hep-ex/9412004\\
\vspace*{3.0cm}
\begin{center}
\begin{Large} {\bf
  On the Kinematic Reconstruction  of  Deep\\
  Inelastic  Scattering at HERA: the $\Sigma$ Method
}\end{Large} \end{center}
\vspace{2.0cm}
\begin{center}
\begin{large}
Ursula Bassler, Gregorio Bernardi  \\
\end{large}
\end{center}
%vspace{0.8cm}
\begin{center}
     Laboratoire de Physique Nucl\'eaire et de Hautes Energies\\
     Universit\'e de Paris VI-VII, 4 Place Jussieu,
         Paris 75230 cedex 05, France\\
\end{center}
\vspace{2.5cm}
\begin{abstract}
%noindent
We review and compare the  reconstruction methods of the inclusive deep
inelastic scattering  variables used at HERA.
We  introduce a  new prescription,
                             the Sigma ($\Sigma$) method, which
             allows to measure the structure function of the proton
$F_2(x,Q^2)$ in a large kinematic domain, and in particular
               in  the low $x$-low
$Q^2$ region, with small systematic errors and small radiative
corrections.  A detailed comparison between
       the       $\Sigma$ method and the            other
methods is shown. Extensions of the $\Sigma$ method are presented.
The effect of QED radiation on the kinematic
reconstruction and on
                   the structure function measurement is discussed.
\end{abstract}
\end{titlepage}

\section{Introduction}
The measurement of the structure functions of the nucleon requires
a precise reconstruction of the deep inelastic scattering (DIS)
                                    kinematics. With the advent of
HERA, the first electron-proton collider ever built, this reconstruction
needs not anymore to rely      on  the scattered lepton only, since the
most important  part of the
hadronic system is visible in the $ 4 \pi$ detectors H1 and
ZEUS. This redundancy  allows for
                              an experimental control of the systematic
errors on the structure function
measurement  if it is based on
 several independent methods to determine the  usual DIS
kinematic variables $x,y$ and $Q^2$:
\begin{equation}
  x = \frac{Q^2} {2 P.q}    \hspace*{4.5cm} y = \frac{P.q} {P.k}
\end{equation}
\begin{equation}
 Q^2= -(k-k')^2 = -q^2      \hspace*{2cm} Q^2 = xys
\end{equation}
with $s$ being the  center of mass energy squared,
 $P,k $  the 4-vectors of the incident proton and lepton
and $k'$ the scattered lepton one.
In this report we  briefly review in section 2 the methods used at HERA
in the first year of operation.
%                              and discuss the possibilities
%of the  kinematic fit.
We introduce in section 3 a
new way   to compute the kinematics,
the $\Sigma$ method, which allows a
precise $F_2$ measurement in the complete kinematical region accessible
at HERA including the   low $x$-low $Q^2$ region.
% although relying
%mainly on the hadronic final state.
We then  make two
rather detailed comparisons of the $\Sigma$, the mixed,
the double-angle  and the electron
methods at low $x$ and at high $Q^2$ (sect.4).
Extensions of  the $\Sigma$ method, including a  prescription
for kinematic fitting are given in section 5.
We  examine  the
effect of  QED radiation
on the $F_2$ measurement (sect.6). Two methods
independent of initial state QED radiation  are      defined
and compared in section 7.
             The main conclusions are summarized in section 8.
\section{Kinematic Reconstruction Methods at HERA}
In the naive quark-parton model, the lepton scatters elastically
against a quark of the proton, and the two body final state is
completely constrained using two variables, if we know the initial
energies labeled $E^e$ and $E^p$ for the electron and proton.
Denoting $E, \theta$ and  $F_q, \gamma_q$  the energy and angle of the
electron and of the struck quark (which gives rise to
the ``current jet"), we can derive
$y$ and $Q^2$ by combining any 2  of them.
For a scattered quark $F_q$ and $\gamma_q$ are
well defined observables, but what is actually observed
after a collision is a set of particles making energy deposition in the
detector. After identification of the scattered electron, we can
define 2 independent quantities: $\Sigma$, defined
as the sum of the scalar quantities $E_h-p_{z,h}$ of each particle
belonging  to the hadronic final state, and $T$
  as its total transverse momentum.
%of the hadronic final state, and its total $\delta$ denoted $T$ and
%$\Sigma$ in the following.
%
\begin{equation}
 \Sigma = \sum_{h} (E_h-p_{z,h})
                                  \hspace*{2cm}
     T   = \sqrt{(\sum_{h} p_{x,h})^2+(\sum_{h} p_{y,h})^2 }
\end{equation}
$E_h,p_{x,h},p_{y,h},p_{z,h}$ are the four-momentum vector components
of  each hadronic final state particle. From energy-momentum
conservation
%with $h$ symbolizing all particles belonging to the hadronic final
%state.
   we can then define an inclusive angle $\gamma$ but also
an inclusive energy $F$ of the hadronic final state which would
be the same as   the angle and
                     energy of the scattered quark in the naive
quark-parton model. The formulae which allow to go from $\Sigma,T$
to $F,\gamma$ are:
\begin{equation}
    \tan\frac{\gamma}{2} = \frac{\Sigma}{ T}   \hspace*{4cm}
   F = \frac{ \Sigma^2 + T^2 } {2 \Sigma} \
%    = \ \frac{\Sigma} {2 \sin^2\frac{\gamma} {2}  }
\end{equation}
 $\Sigma$ is by construction minimally affected by  the
losses in the forward direction\footnote{The positive $z$
axis is defined at HERA as the incident proton beam direction. The
electron polar angle is thus large ($\sim 170^o$) for small  scattering
angle, i.e. small $Q^2$.}
($= 2 E_i \sin^2\frac{\theta_i}{2}$ for every lossed particle $i$)
due to the beam pipe hole in which
 the target jet and the initial state gluon radiation tend to
disappear. $T$ covers the other spatial dimensions,
but it is more sensitive to forward losses ($=  E_i \sin{\theta_i}$),
thus to improve kinematic reconstruction we should try to avoid using
it. We can use $\gamma$ instead, which carries the $T$ information
and  is better measured since in the ratio $\Sigma/T$
                              the energy uncertainties cancel
                 and the effect of the losses in the forward
beam pipe is  diminished.
In conclusion  the optimal four  ``detector oriented" variables
to characterize deep inelastic scattering at HERA are
$[E,\theta,\Sigma,\gamma]$.

%subsection{Basic Methods}
If we want to determine $x$ and $Q^2$ from 2  variables
out of these 4,
%            coming from the electron and the hadronic system,
                                                 we are
left with 6 combinations  of which only 3 have been shown to
                                      be  usable
for structure function measurement (the formulae are given in the
appendix):
\begin{itemize}
\item
                 the electron-only ($e$) method which uses
                                           $E$ and $\theta$
is experimentally  simple and  very precise
in the high $y$ ($> 0.2$) region. It nevertheless degrades rapidly
with decreasing $y$~\cite{JoJo} due to the $1/y$ term in the
      error propagation\footnote{We define $A \oplus B \equiv \sqrt{A^2
+B^2}$} of $y_e$:
\begin{equation}
\frac{\delta y}{y} = \frac{1-y}{y} \ (\frac{\delta E}{E} \oplus
                                  \frac{\delta \theta}{\tan{\theta/2}})
\end{equation}
\item
     the hadrons-only ($h$) method~\cite{jb} provides  a rather precise
  measurement of $y$ at low and medium $y$ ($y<0.2$), which degrades
at high $y$. It gives
    a rather poor measurement of $Q^2$ due to the
loss of hadrons in the beam pipe but  it is  the only possible
                                                      method for
``charged current" events.
\item
the double-angle (DA) method~\cite{ben1, hoe} uses the angle of the
electron $\theta$ and the inclusive angle of the hadronic final state
$\gamma$. Assuming an homogeneous energy measurement over the full
solid angle, we can deduce from eq.4 that the DA method is independent
of the absolute energy calibration.
%          From eq.4 we can remark that $\gamma$ is independent
% of the absolute energy scale.
                              This method is precise at high $Q^2$
($Q^2> 100 GeV^2$) but has large resolutions at
low $x$-low $Q^2$~\cite{gbwh}, as will also be
                    illustrated in section 4.
\end{itemize}
At HERA a fourth method has been used which combines $y$ measured
from the hadrons and
$Q^2$ from the electron, which is the so-called mixed ($m$)
method~\cite{maxu}:
\begin{equation}
 x_{m} = \frac{Q^2_e} {s~y_h} \hspace*{1.3cm}\mbox{with}\hspace*{1.3cm}
 y_{h} = \frac{\Sigma} {2 E^e}
\end{equation}
The mixed method allows to extend towards lower $y$
                                  the $F_2$ measurement done
with the DA method~\cite{gbwh} or
%at high $y$
with the $e$ method~\cite{gbwh,H193}.
Note that in the mixed method more than 2 independent variables are
used, i.e. $E,\theta$ and $\Sigma$, justly avoiding $T$,
and
we will see in the next section that these 3 variables can be combined
in a more efficient way.
The basic  methods presented above
        become imprecise in some places of the kinematic
 domain.  To have a single method covering the complete kinematic
region measurable at HERA  we can either make use of the kinematic
                                                   fitting
procedure (see sect.5) or explore further combinations of the 4
variables $[E,\theta,\Sigma,\gamma]$ (sect.3 and 7).

\section{The $\Sigma$ method}
With 3 variables, we can determine   $y$ and $Q^2$ independently of
initial state QED radiation (ISR) by reconstructing
   the incident electron energy.
{}From   energy-momentum conservation, the measured
            quantity $\Delta$ is equal to two times the electron
beam  energy, if no particle escape detection:
\begin{equation}
                                     \Delta \equiv
        \Sigma + E (1-\cos{\theta})
= 2 E^e
\end{equation}
We can then obtain from  $y_{h}$ a new
expression  $y_{\Sigma}$~\cite{bas}  which gives $y$ at the
hard interaction vertex even if an
               ISR photon is emitted.
\begin{equation}
   y_{\Sigma} = \frac{\Sigma}{ \Sigma + E (1-\cos{\theta})}
\end{equation}
$y_{\Sigma}$
   has another  important characteristic:
 at high $y$, i.e in the most interesting region at low $Q^2$,
 $\Sigma$ dominates over $E (1-\cos{\theta})$, and the
fluctuations and errors
             on $\Sigma$ start cancelling between numerator and
denominator, allowing an optimal $y$ reconstruction when using the
                                                         hadrons.
Writing  the error propagation on $y_{\Sigma}$,
\begin{equation}
\frac{\delta y}{y}= (1-y) \ (\frac{\delta \Sigma}{\Sigma} \oplus
                             \frac{\delta E}{E} \oplus
                   \frac{\delta \theta}{\tan{\theta/2}})
\end{equation}
we can observe the following differences with $y_e$ (eq.5)
and $y_h$ ($\delta y_h/y_h=\delta \Sigma/\Sigma$): \\
i) At low $y$
   there is no divergence of $\delta y/y$  for $y_{\Sigma}$, which
%                                              $y_{\Sigma}$
      has    practically the same behaviour as       $y_h$
since the $\delta \Sigma / \Sigma$ term is experimentally bigger
than the two terms coming from the electron measurement.\\
ii) at high $y$, the error on $y$ decreases as $(1-y)$ for $y_{\Sigma}$
but is still dominated experimentally by the error on $\Sigma$.
The  $\delta\Sigma/\Sigma$ term is partly compensated by
the $(1-y)$ term contrarily to the $y_h$ case.

%  After this improvement at high $y$, the correlation with $y_e$
%becomes stronger and any significant deviation from this correlation
%signs either an initial state radiation event, since it influences
%$y_e$, or a faked electron from an photoproduction background event,
%whose rate increase dramatically at low $E$.
In fig.1 is
shown the behaviour at low $x$-low $Q^2$ (see sect.4.1 for the
description of the sample of Monte Carlo events used here)
               of the correlation  $y_{\Sigma}/y_e$
compared to $y_h/y_e$ and to $y_{DA}/y_e$. These predicted behaviours
have been reproduced by the H1 collaboration for the $F_2$ analysis on
        the  1993  data~\cite{H194}.  We can observe the
expected improvement of $y_{\Sigma}$ at high $y$, despite the
worsening of $y_h$ due to the low energy of the hadronic final
state\footnote{This effect is amplified in the H1 detector by the poor
hadronic calorimeter coverage in the backward direction, in which the
hadronic final state particles tend to go at high $y$. We can remark
that $y_{\Sigma}$ is relatively  insensitive to this peculiarity.}.
In this plot $y=0.5$ corresponds in average to $x \sim 2.10^{-4}$.
\begin{figure}[htbp]\centering
\epsfig{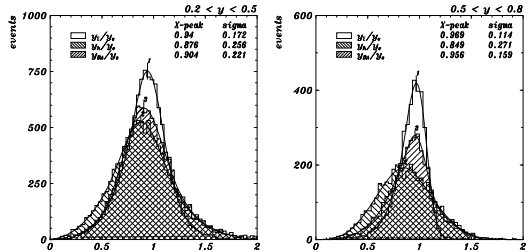}
\caption[]{\label{yel}
\sl Monte Carlo comparison of $y$ reconstruction for several hadronic
methods
w.r.t. to $y_e$, at high (0.2-0.5) and very high $y$ (0.5-0.8).
Both bias and resolution improve with $y$ for the $\Sigma$ method
although these become worse for the hadrons-only method. $y_{DA}$
has larger biases and resolutions than $y_{\Sigma}$, as can be
read from the fit results  written inside the figure.}
\end{figure}

{}From $E,\theta$ and $\Sigma$ we determine $Q^2_{\Sigma}$, also
independent of ISR  when assuming that an initial state radiated photon
which is colinear to the beam line do not carry transverse momentum.
\begin{equation}
   Q^2_{\Sigma} = \frac{E^2 \sin^2{\theta}}{ 1-y_{\Sigma}}
\end{equation}
{}From $Q^2=xys$
we can obtain $x_{\Sigma}$ in two ways: \\
                             i) taking the ``true" $s$ at the
hard interaction vertex, as will be dicussed in section 7.
                              ii) taking $s$ as  computed from the
beam energies ($s=4E^eE^P$):
%Here $x_{\Sigma}$ is defined as
%$x_{\Sigma}$ using the nominal $s$:
\begin{equation}
 x_{\Sigma} = \frac{Q^2_{\Sigma}}{s y_{\Sigma}}
            = \frac{E^2 \sin^2 \theta} {s~y_{\Sigma}(1-y_{\Sigma})}
\end{equation}
Thus we have  defined a new method, which has similar characteristics
as  the mixed method at low and medium $y$, but a higher precision
at  high $y$ allowing to  cross-check  the $e$ method
in this region. It must be stressed that this improvement does not
come from the precision of the electron measurement, on which
on the contrary relies the precision of the $e$ method. These
methods then provide two measurements essentially independent.
Finally $y_{\Sigma}$ and $Q^2_{\Sigma}$ being ISR independent, the
radiative corrections to the $\Sigma$ method are smaller than to the
other methods  (see sect.6).

\section{Comparison of Electron, DA, Mixed and $\Sigma$ methods}

We study in this section the resolutions and biases of the
$x$ and $Q^2$ variables as reconstructed by the different methods.
The Monte Carlo event sample on which this study is based has been
generated according to cross-section above $5~GeV^2$,
with the DJANGO~\cite{django} program which relies on
HERACLES~\cite{heracles} for the treatment of QED radiation, and
for the simulation of the hadronic final state on the
color dipole model (CDM) as implemented in ARIADNE~\cite{cdm}. The CDM
is known to give the best description
of the energy flow on the HERA data~\cite{h1flow, zeflow}. The
MRSH~\cite{mrsh} parton densities parametrizations, which include
the constraints from the 1992 HERA data  were used. The generated
events were then processed through the H1 analysis chain, which
includes a detailed simulation based on GEANT~\cite{geant} of the
H1 detector~\cite{h1dete}. Thus the results presented here after
include in a realistic way the effect of inhomogeneities or small
miscalibrations encountered in a  real  detector. The conclusions
that we will draw are nevertheless general, since the characteristics
of the H1 and ZEUS detectors are rather similar after the
major detector effects are corrected by the reconstruction program.
To make a meaningful comparison, all methods are applied on the same
events using the same H1KINE~\cite{h1kine} software package which is
part of the H1 reconstruction program. One of its characteristics is to
use a combination of track momentum and calorimetric energy
in the central part of the detector ($25^o<\theta<155^o$)
to improve the precision on the hadronic final state measurement and in
particular to render it sensitive to the low energy hadrons
($E_h<0.5~GeV$) present in large number
at low $Q^2$. This treatment is applied consistently to all
reconstruction methods.

The structure function analysis is generally done in
bins of $x$ and $Q^2$, thus we will study here the following
distributions\footnote{When $x,y$ or $Q^2$ have not the
subscript which corresponds to a given method, they are assumed to be
the true i.e. generated values at the hard interaction vertex.}:
                                 $x_{method}/x$ and $Q^2_{method}/Q^2$.
No bias on the reconstructed variable is observed when the
distribution peaks at 1.
                       In order to conserve a good proportion of events
originating from a given bin in that bin (defined as the ``purity'' of
the bin)
we need to have small biases and good resolutions on $x$ and $Q^2$.
The better the resolution the finer binning we can reach, assuming
we have enough events, and thus the more subtle effects in the $F_2$
evolution along $x$ or $Q^2$ can be observed.
This is particularly true at low $x$ where the number of events
collected in the first three years of operation is large enough
to make the finest binning  allowed by the detector resolutions.

We will focus successively on two very different kinematic regions,
for obvious physical reasons: the low $x$ domain ($x < 10^{-3}$)
which at HERA is accessible only below $\sim  50~GeV^2$, and the
high $Q^2$ domain  which we will define as $Q^2 > 200~GeV^2$. In both
cases,
 the events must also satisfy the following
conditions: $\theta< 173^o$, $E>8~GeV$, $\Delta > 30~GeV$, this last
condition being helpful to reduce radiative effects and, on data,
the background to the deep inelastic interactions.
However
for the distributions shown in the next two
                                  sections, only
                                      events without  emitted photons
from the electron line are  used, in order not to mix  the smearing
due to the reconstruction  method with the one coming from the radiation
effect.
\subsection{Comparison at Low $x$}

\begin{figure}[htbp]
\epsfig{file=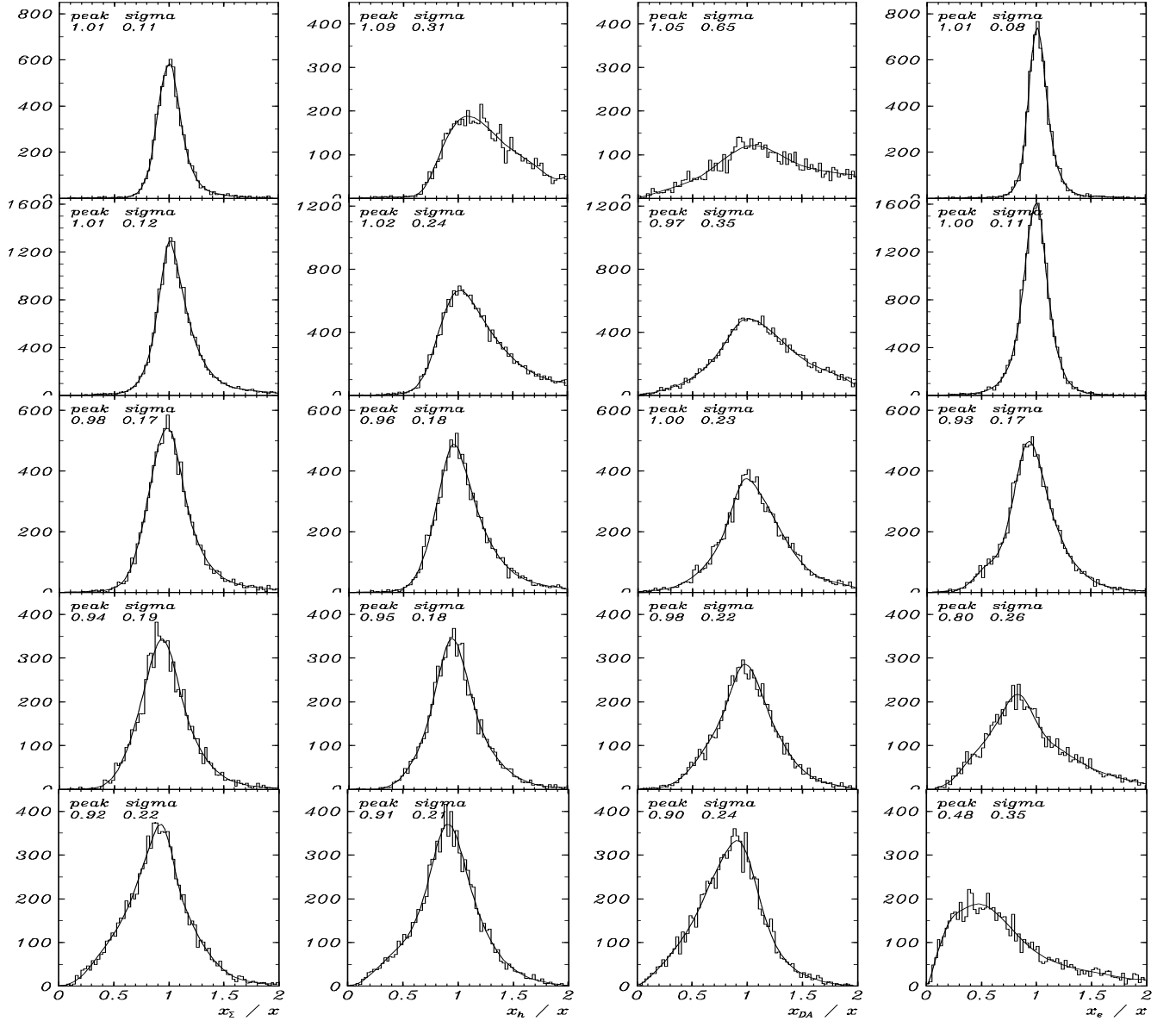,width=16cm,height=18cm,angle=90.}
\caption[]{\label{res1}
\sl Comparison $x_{method} / x$ at low $Q^2$
    ($Q^2 > 7~GeV^2$) for the $\Sigma $, mixed , DA and $e$ methods.
                                 From top to bottom,
each row represent a bin in $y$:
                                                    very high
(0.5-0.8), high (0.2-0.5), medium (0.1-0.2), low (0.05-0.1), very low
(0.01-0.05).                                                     }
\end{figure}
\begin{figure}[htbp]
\epsfig{file=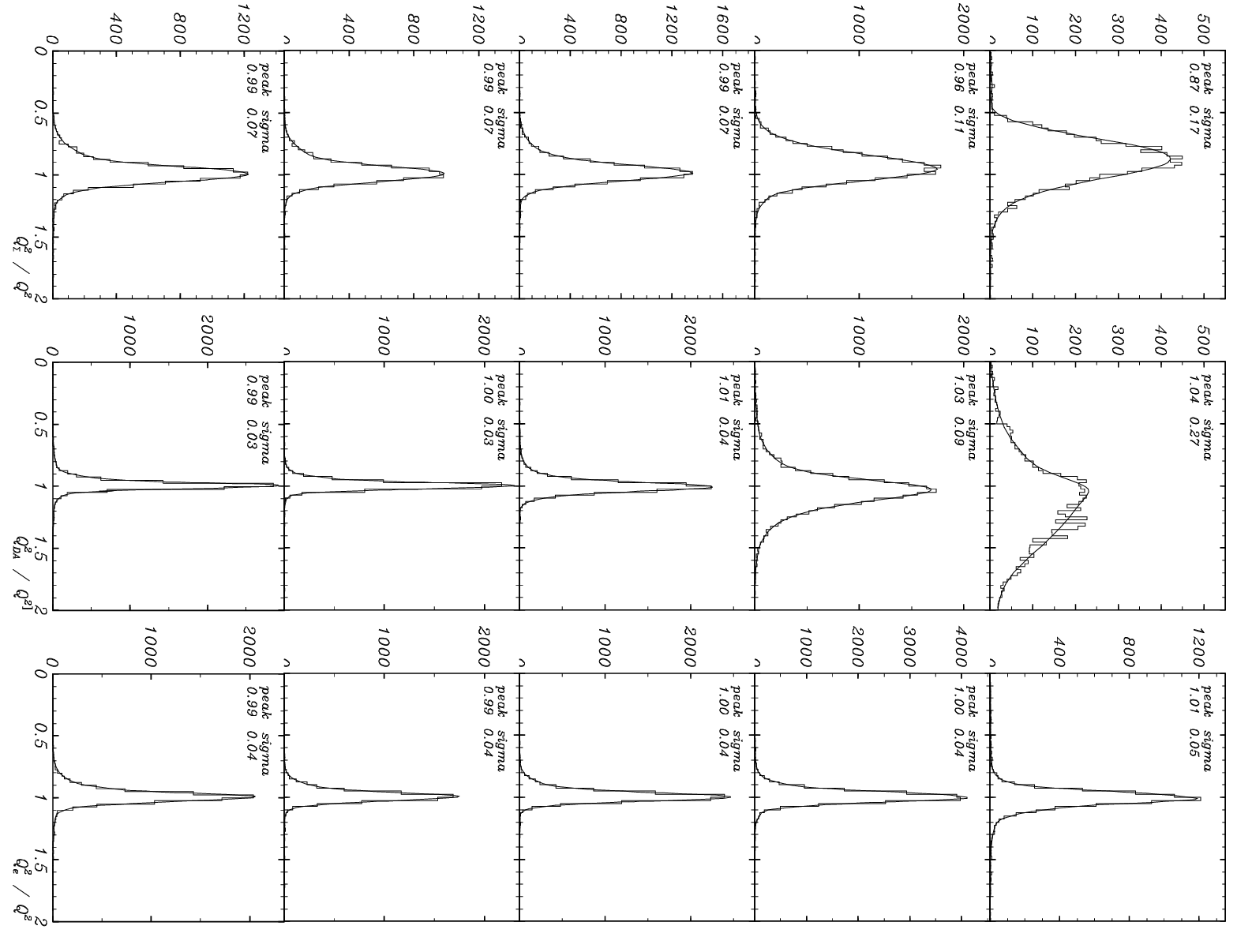,width=16cm,height=18cm,angle=90.}
\caption[]{\label{res2}
\sl Comparison $Q^2_{method} / Q^2$ at low $Q^2$
    ($Q^2 > 7~GeV^2$) for the $\Sigma $ , DA and $e$ methods.
                                 From top to bottom,
each row represent a bin in $y$:
                                                    very high
(0.5-0.8), high (0.2-0.5), medium (0.1-0.2), low (0.05-0.1), very low
(0.01-0.05).                                                     }
\end{figure}

For this study we ask: $Q^2 > 7~GeV^2$.
The results are shown in figs.2 and 3 for $x$ and $Q^2$
respectively, with one column per method and 5 bins in $y$.
We distinguish very high (0.5-0.8), high (0.2-0.5), medium (0.1-0.2),
low (0.05-0.1) and very low $y$ (0.01-0.05), reminding
that the lowest $x$ are at highest $y$ and viceversa.
{}From these distributions  we can draw the following
 conclusions which are valid in the low $Q^2$ region:
\begin{itemize}
\item
The $\Sigma$ method has, as expected, a  much more precise
$x$ reconstruction than the mixed
method at high and very high $y$. The behaviour of $x_{\Sigma}$ and
 $x_m$ is almost identical at low and medium $y$.
\item
$x_{\Sigma}$ is the only $x$ which is reconstructed with small
biases ($<10\%$) and precise resolutions ($< 20\%$)
over the full $y$ range
(0.01-0.8). $x_{DA}$ can compete at low and medium $y$, but becomes
very imprecise when $y$   rises. $x_e$ is slightly more
precise than $x_{\Sigma}$ at high $y$, then rapidly becomes
imprecise with decreasing $y$.
\item
$Q^2_e$ has always very small biases and optimal resolutions
($\sim 4\%$). $Q^2_{\Sigma}$ has no bias and
                               good resolution ($\sim 7\%$)
at low and medium $y$, but becomes more imprecise with increasing
$y$.  $Q^2_{DA}$ has the same tendency, but in a more extreme way:
better resolution at low $y$~($\sim 3\%$)
                      and  poorer resolution at very high $y$.
\end{itemize}
In conclusion, for the low $x$ physics , i.e. at low $Q^2$, the
$\Sigma$ method allows to make a complete measurement in the
full            kinematic range. The $e$  method remains
the most precise method at high $y$, but now it can be compared to
a hadronic method in the low $x$ domain.
The  DA  method is clearly disfavoured  for
low $x$  physics at low $Q^2$. Finally
the mixed method can be advantageously replaced by the $\Sigma$ method.

\subsection{Comparison at High $Q^2$}
%For this study, we request $Q^2 > 200~GeV^2$ and the corresponding
%distributions are shown in figs.4 and 5. We can draw the following
%conclusions:
For this study, we request $Q^2 > 200~GeV^2$ and the $x_{method}/x$
distributions are shown in fig.4. The corresponding $Q^2$ distributions
have all a gaussian shape,
                    so we just give their mean value and standard
deviation in table 1.
\begin{table}[htbp] \centering
\begin{tabular}{|c|c|c|c|}
\hline
               &$Q^2_{\Sigma}/Q^2$ &  $Q^2_{DA}/Q^2$ & $Q^2_{e}/Q^2$ \\
\hline
$0.5< y< 0.8$&$0.89~\sigma=0.16$&$0.99~\sigma=0.07$&$0.99~\sigma=0.04$\\
$0.2< y< 0.5$&$0.98~\sigma=0.08$&$0.99~\sigma=0.03$&$0.99~\sigma=0.03$\\
$0.1< y< 0.2$&$1.01~\sigma=0.06$&$0.98~\sigma=0.03$&$1.00~\sigma=0.03$\\
$0.05<y<0.10$&$1.01~\sigma=0.05$&$0.98~\sigma=0.02$&$0.99~\sigma=0.03$\\
$0.01<y<0.05$&$1.02~\sigma=0.05$&$0.98~\sigma=0.02$&$1.01~\sigma=0.03$\\
\hline
\end{tabular}
\caption[]{\label{table1}
\sl Comparison $Q^2_{method} / Q^2$ at high $Q^2$
    ($Q^2 > 200~GeV^2$) for the $\Sigma $ , DA and $e$ methods.}
\end{table}

\noindent
{}From fig.4 and tab.1 we can draw the following
conclusions:
\begin{itemize}
\item
The general behaviour of the ${\Sigma}$, the mixed
and  the $e$ method have not
changed, both in $x$ and $Q^2$. A slight improvement in precision
is visible at higher $Q^2$, which is clearly understood
by a more precise energy measurement of the electron and of the hadrons.
\item
The DA method becomes more precise and improves at higher $Q^2$:    \\
- In $x$ it has better resolutions
than the $\Sigma$ method for $y$ between 0.05 and 0.3. It is more
affected by losses in the beam pipe at very low $y$, and not as
precise as the $e$ or $\Sigma$ methods
at high $y$, but this last feature
will disappear at very high $Q^2$ (above 2000~$GeV^2$).\\
- In $Q^2$ it is very precise at medium and low $y$,
and display a similar behaviour as $Q^2_{\Sigma}$ at higher $y$,
but without biases and better resolutions. The $Q^2_{DA}$ resolution
remains below $10\%$ over the full $y$ range.
\end{itemize}
In conclusion, it is also
possible to make a very precise kinematics reconstruction at high
$Q^2$, although this precision is not as important as at low $Q^2$ due
to the fast decrease of the DIS cross-section with $Q^2$, which
will always prevent at HERA  a very fine ``binning".

\begin{figure}[htbp]
\epsfig{file=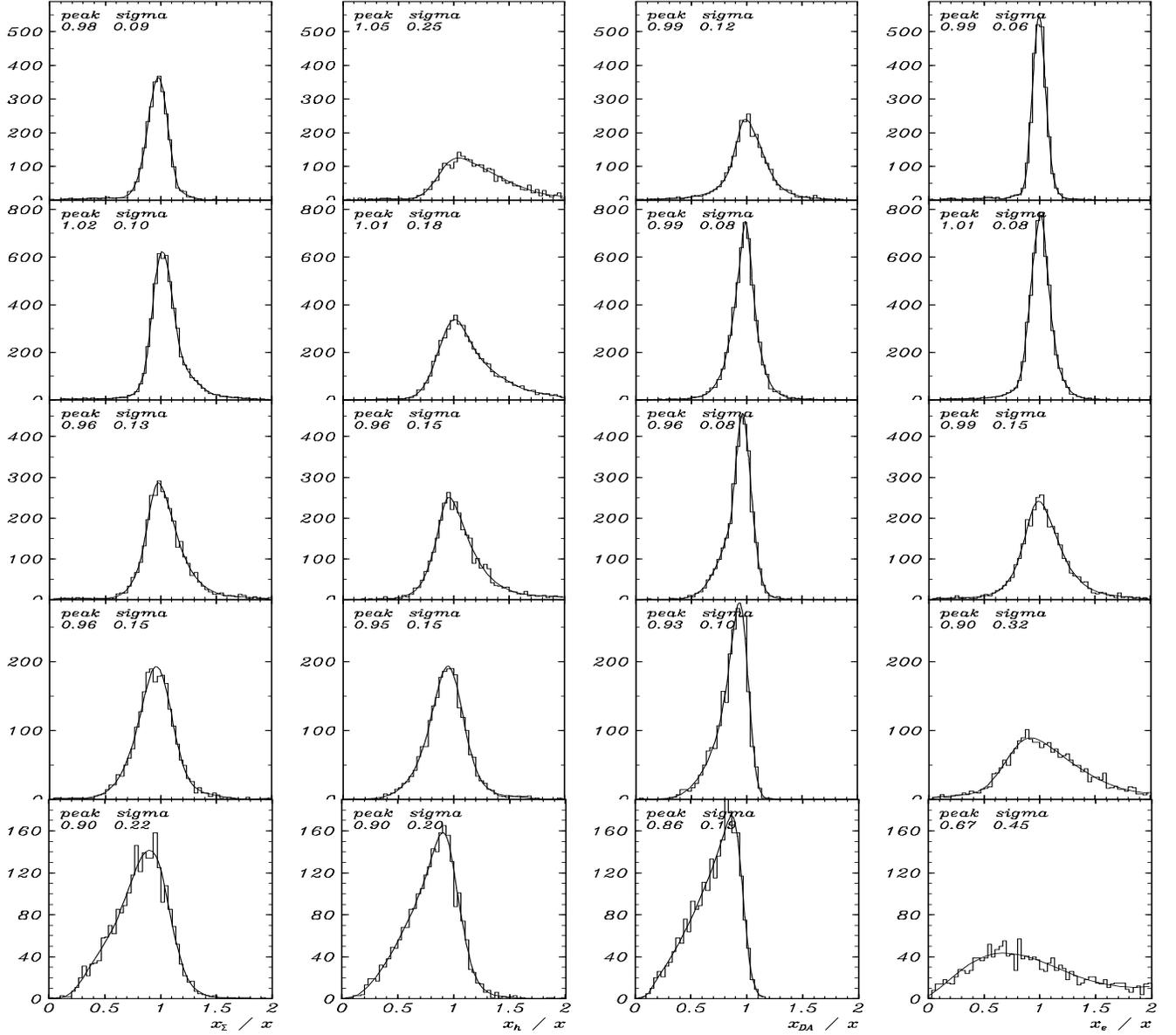,width=16cm,height=18cm,angle=90.}
\caption[]{\label{res3}
\sl Comparison $x_{method} / x$ at high $Q^2$
    ($Q^2 > 200~GeV^2$) for the $\Sigma $, mixed , DA and $e$ methods.
                                 From top to bottom,
each row represent a bin in $y$:
                                                    very high
(0.5-0.8), high (0.2-0.5), medium (0.1-0.2), low (0.05-0.1), very low
(0.01-0.05).                                                     }
\end{figure}
%
%\begin{figure}[htbp]
%\epsfig{file=res4.eps,width=16cm,height=18cm,angle=90.}
%\caption[]{\label{res4}
%\sl Comparison $Q^2_{method} / Q^2$ at high $Q^2$
%    ($Q^2 > 200~GeV^2$) for the $\Sigma $ , DA and $e$ methods.
%                                 From top to bottom,
%each row represent a bin in $y$:
%                                                    very high
%(0.5-0.8), high (0.2-0.5), medium (0.1-0.2), low (0.05-0.1), very low
%(0.01-0.05).                                                     }
%\end{figure}

\subsection{Effect of QED Radiation on Kinematic Reconstruction}
In this section we study the QED radiation effect on the kinematic.
We take the same selection criterias than in sect.4.1, but requesting
now the presence of a radiated photon having an energy  such that
$1.5~GeV < E^{\gamma} < 12.5~GeV$, the upper bound coming from the
$\Delta$ cut.
%This sample is just complementary to the one described in 4.2 .
The true $x,y$ and $Q^2$ are defined at the hard interaction vertex
                                    taking into account
the ISR or final state radiation (FSR) effects:
For the ISR, these variables
      are computed assuming an incident electron  energy equal to
$E^e-E^{\gamma}$.
For the FSR, the photon is summed with the electron. This is
experimentally happening in most of the events since the colinear
photon is merged in the cluster of energy deposited by the electron
in the calorimeter.

Since the results are in
general very   similar to the ones obtained excluding the QED
effects, we can just summarize this
comparison on  the biases and resolutions of the distributions with
and without radiation. \\
i) the $Q^2$      reconstruction is essentially not affected by
radiation in all the methods studied. The additional biases and
enlargement of resolution are always smaller than $2\%$. \\
ii) the $x$  reconstruction in the hadronic methods ($m$,DA,$\Sigma$)
is also very stable against radiation, with changes of the order
of $1$  to $3\%$.\\
iii) the most affected variable  is  $x_e$  when going at low $y$.
  The additional bias due to radiation on $x_e/x$ is
in the 5 bins, from very high to very low $y$: $-1\%, -5\%, -10\%,
                 -20\%, -35\%$.

%\begin{table}[htbp]
%\caption[]{\label{resno1}
%\sl Comparison $x_{method} / x$ at low $Q^2$
%    ($Q^2 > 7~GeV^2$) for $\Sigma-$, m, DA and $e$ method. Radiative
%    events.}
%\end{table}
%\begin{table}[htbp]
%\caption[]{\label{resno2}
%\sl Comparison $Q^2_{method} / Q^2$ at low $Q^2$
%    ($Q^2 > 7~GeV^2$) for $\Sigma-$, m, DA and $e$ method. Radiative
%    events.}
%\end{table}

\section{Beyond the $\Sigma$ Method}
\subsection{The $e\Sigma$ Method}
{}From the  previous section, we can conclude that the $\Sigma$ method
can replace the mixed method and  provide, with a single prescription,
a precise $F_2$
measurement over the complete range covered by at least one the
4 basic methods~\cite{H194}.
%         The 3 methods ($e,\Sigma$,DA) can be used for structure
%function measurements and compared to estimate the systematic
%errors~\cite{H194, ZEUS94}.
It is however clear from figs.2-4 than the two most stable
reconstruction for $x$ and $Q^2$ all over the kinematic plane
are $x_{\Sigma}$ and $Q^2_e$.
We can then simply define a so-called $e\Sigma$ method,
in which the $Q^2$ is determined from the electron quantities,
and $x$ is reconstructed according to the $\Sigma$ prescription:
\begin{equation}
 x_{e\Sigma} = x_{\Sigma} \hspace{2cm}
 Q^{2}_{e\Sigma} = Q^{2}_e \hspace{2cm}
 y_{e\Sigma} = \frac{2~E^e~\Sigma}{(\Sigma+E(1-\cos\theta))^2}
%            = \frac{2~E^e}{\Sigma+1-\cos\theta} y_{\Sigma}
\end{equation}
The improvement brought by this
                        method can also be observed on $y_{e\Sigma}$,
by comparing its correlation
                          $y_{e\Sigma}/y_e$ to the ones shown in fig.1,
%about $y_{\Sigma},y_h,y_{DA}$,
                               as can be also  read in table 2.
\begin{table}[htbp] \centering
\begin{tabular}{|c|c|c|c|}
\hline
               &   $0.2 < y < 0.5$        &         $0.5 < y < 0.5$  \\
\hline
$y_h/y_e$      & $ 0.88~\sigma=0.26$      &  $ 0.85~\sigma=0.27$ \\
$y_{DA}/y_e$   & $ 0.90~\sigma=0.22$      &  $ 0.96~\sigma=0.16$ \\
$y_\Sigma/y_e$ & $ 0.94~\sigma=0.17$      &  $ 0.97~\sigma=0.11$ \\
$y_{e\Sigma}/y_e$  & $ 1.00~\sigma=0.06$      &  $ 1.01~\sigma=0.03$ \\
\hline
$y_{\Sigma} /y $   & $ 0.95~\sigma=0.15$      &  $ 0.95~\sigma=0.11$ \\
$y_{e\Sigma}/y $   & $ 0.99~\sigma=0.09$      &  $ 0.99~\sigma=0.06$ \\
$y_{e}      /y $   & $ 0.99~\sigma=0.06$      &  $ 0.99~\sigma=0.03$ \\
\hline
\end{tabular}
\caption[]{\label{eSigma}
\sl Comparison of $y$ reconstruction quality, for the
    $\Sigma$, mixed, DA, $e$ and $e\Sigma$ method.}
\end{table}
The correlation with $y_e$ is excellent at high $y$, in fact higher
than its  correlations to the  generated $y$. We thus have a method
which   optimizes the reconstruction quality,
but has the disadvantage of becoming correlated to the $e$ method
at high $y$ and also to  enlarge the
small influence of radiative correction displayed by the $\Sigma$
method as we shall see in  section 6.
\subsection{Kinematic Fitting}
The ultimate improvement for the $(x,Q^2)$ determination
will come from a kinematical fitting procedure which use the
full information in an optimal way.
The weights on the redundant information coming from the electron
and the hadronic final state have to  be assigned depending
on the kinematic of the event itself, and need, for a precise
determination, a deep knowledge of the detector response.
So far the fit has been performed using the conservation of
$\Delta$ and of the transverse momentum~\cite{hoe,chaves,julian}
(for simplicity, we do not consider here the staightforward modification
needed to treat the ISR case).
This is equivalent to say that the fit has been performed using
$[E,\theta,\Sigma,T]$ as measured variables and in fact
using only $[E,\theta,\Sigma]$ since $T$ is badly measured.
                                                  It is then
a posteriori clear  why the fit behaves only slightly better than
                                       the $e$ method at high $y$,
the $\Sigma$ method at medium $y$, and the $\Sigma$ or $m$ method
at low $y$.

To go beyond this ``3-variables fit"  we need to
  include  in the fitting procedure
    the  information contained in the hadronic transverse
momentum. This can be achieved by
    replacing (cf sect.2) $T$  by the hadronic angle $\gamma$.
We can then fit
$y$ and $Q^2$    according to the following equations
\begin{equation}
  E (1-\cos \theta) = 2 E^e (1-y) \hspace*{2cm}
  E \sin \theta = \sqrt{Q^2 (1-y)}
\end{equation}
\begin{equation}
  \Sigma = 2 E^e y \hspace*{4cm}
  \tan\frac{\gamma}{ 2} = 2 E^e \sqrt{\frac{y^2}{Q^2 (1-y)}}
\end{equation}
hence the constraint equations  can be written as
\begin{equation}
  \Sigma+E(1-\cos\theta) = 2 E^e  \hspace*{2cm}
  \tan\frac{\gamma}{ 2} = \frac{2 E^e -E(1-\cos\theta)}{E\sin\theta }
\end{equation}
Assuming that  the error covariance matrices has  been determined
for eqs.13-14, this fit
                       will then provide everywhere a better result
than any of the $e,h,m,DA,\Sigma$ methods, i.e. it will
 also have    the properties of the
                                  DA method like its high precision
at large $Q^2$.

A complete
  study of the kinematic fitting  is beyond the scope  of this
paper but we can conclude that there are at least
     two ways to become more precise than the basic or
 $\Sigma$ methods,                       which  both  have the
drawback that the measurement they can provide
       cannot be cross-checked a posteriori
       in a simple and independent way.

\section{Comparison of QED Radiative Corrections}

\begin{figure}[htbp]\centering
\epsfig{file=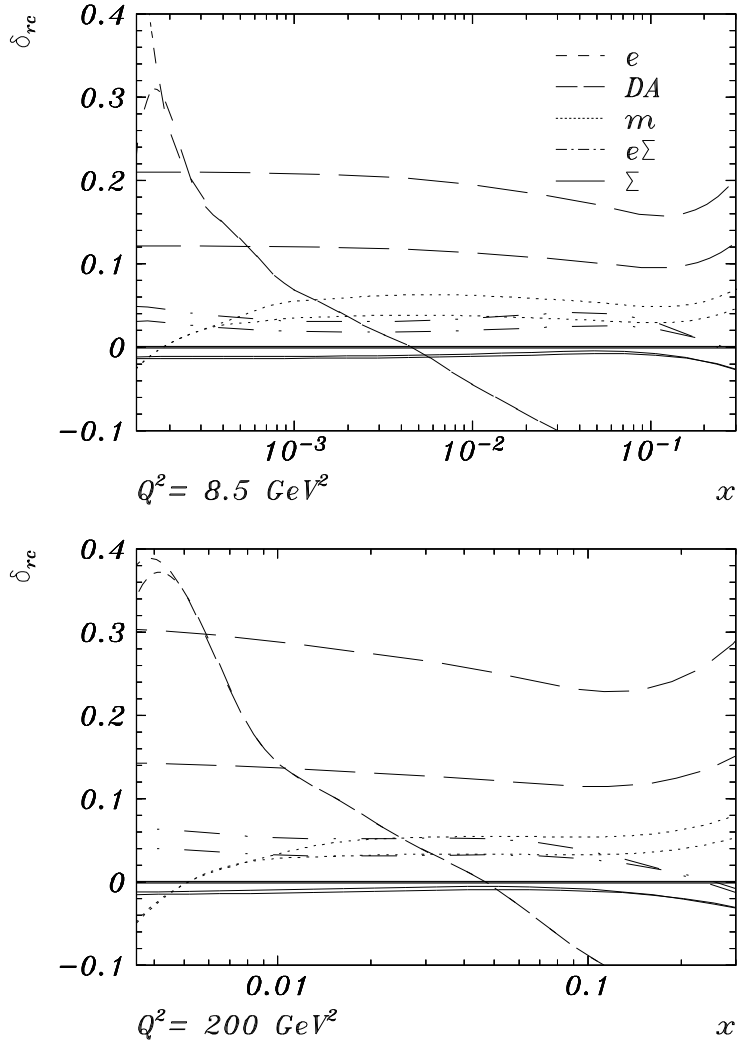,width=14cm,angle=90.}
\caption[]{\label{rc1}
\sl Radiative corrections (excluding vacuum polarisation correction)
    as a function of $x$ in two       bins of $Q^2$
    for the $\Sigma$, $e\Sigma$, DA, $e$ and mixed method,
    requesting two minimal scattered electron energy: $E > 4~GeV$
    (upper curve of each method)
    and $E> 8~GeV$ (lower curve).}
%In both plots, the lowest $x$
%    on the abscissa correspond to $y \simeq 0.7 $ }
\end{figure}
%Depending on the reconstruction method, the size of the QED
%radiative correction (r.c.) which has to be applied to the measured
%cross-section to obtain $F_2$
%can be very different. For the $e$ method for instance the r.c.
%are reaching relative values
%up to  $\sim 50\%$ in the kinematic region accessible at HERA.
%The one-photon exchange differential cross section is related to $F_2$
%via the following equation in which
%$\delta_{rc}(x,Q^2)$ quantifies the relative  size  of the r.c.
The size  $\delta_{rc}(x,Q^2)$ of the QED radiative correction (r.c.)
which has to be applied to the measured differential cross-section to
obtain $F_2$   depends on the
method used, since the cross-section is determined in bins of
reconstructed $x$ and $Q^2$. We have
%            since the  kinematical variables are reconstructed
%differently and do not always
%coincide with the ones
%      defined at the hard interaction vertex
%      which  appear in the right part of the following equation
\begin{equation}
\left(\frac{d^2\sigma}{dx dQ^2}\right)
\left(\frac{1}{1+\delta_{rc}}  \right)
                          =\frac{2\pi\alpha^2}{xQ^4}
                               (2-2y+\frac{y^2}{1+R}) F_2(x,Q^2)
\label{dsigma}
\end{equation}
where $R\equiv\sigma_L/\sigma_T$ is the ratio of the cross-section
of longitudinally and transversely polarised photons.
The r.c. due to the virtual graphs, which are dominated
by the fermion loops in the self-energy of the photon
(vacuum polarization)  can be included in a ``running"
$\alpha=\alpha(Q^2)$ in this equation~\cite{spies}. These corrections
are a function of $Q^2$ only,
ranging from 5 to 13 $\%$, for $Q^2$ between 5 and $10^4~GeV^2$.
The remaining $\delta_{rc}$
               is dominated by the ISR from the lepton line
and is thus a sensitive probe to distinguish the r.c. behaviour for the
different reconstruction methods.

The radiative corrections for the
$e,m$,DA,$\Sigma$ and $e\Sigma$ methods
were obtained  with the HELIOS
program\cite{blu} in which they  are calculated in the Leading Log
Approximation up to $\cal O$$(\alpha^2)$, and cross-checked with the
HERACLES program as implemented in DJANGO~\cite{django} for the
$e$ and $\Sigma$ methods~\cite{uwe}.

%In the following we will separate
%the contributions to $\delta_{rc}$ from the vacuum
%polarisation graphs, (3 to 10\%, depending on $Q^2$ only),
%and from the $1^{st}$ and the $2^{nd}$ order QED radiation
%contributions. The r.c. were obtained

In fig.\ref{rc1} is shown as a function of $x$ in 2
representative bins of $Q^2$
(8.5 and  200 $GeV^2$) the size of the radiative corrections
for every method,  without cut on  $\Delta$.
%%%%%%%%%%%%%%% excluding  the vacuum polarization correction (V.P.C.).
We nevertheless take into account the minimal electron energy required
for DIS identification. Two typical values  have been used,
$E = 4~GeV$ and $E = 8~GeV$, so two curves for each method
are displayed. The variation of the r.c. for the 2 minimum energies can
reach  large values ($\delta_{rc}^{DA}\simeq 16 \%$) for the DA
or the $e$ method, while it  rather small for the $m$ and $e\Sigma$
($\delta_{rc}^{m,e\Sigma} \simeq 2  \%$)  and almost null  for the
 $\Sigma$ method.
In the $F_2$ measurable region ($0.01 < y <0.8 $)
there is no dependence on $x$ or $Q^2$ neither
for $\delta_{rc}^{\Sigma}$ nor,
at the $2\%$ level, for $\delta_{rc}^{e\Sigma}$, contrarily to
$\delta_{rc}^{m}$ and $\delta_{rc}^{e}$ (dependence in $x$) or
$\delta_{rc}^{DA}$ (dependence in $Q^2$).
We can also remark that using the nominal $s$ in the $\Sigma$
method has a very small effect on the radiative corrections,
since $\delta_{rc}^{\Sigma} \sim -1.5\%$.

The radiative corrections are therefore experimentally very difficult
to control at the $1\%$ level for  the $e$ and $DA$ methods without
introducing a cut on  $\Delta$ of the events e.g. $\Delta >
30~GeV$ which reduces the ISR effects. After such a cut all ``hadronic"
methods have r.c. displaying an almost constant  behaviour in $x$ and
$Q^2$, and amounting at 8.5 $GeV^2$ to $\delta_{rc}^{tot}=\delta_{rc}+
\delta_{rc}^{vac.pol.} \simeq$ +2, +5, +6, +9 $\%$
for the $\Sigma, e\Sigma,
m, $ DA methods respectively. There is still a dependence
on $x$ for $\delta_{rc}^e$, which stops growing for $y$ values
above $z=       \Delta /  2E^e $.

%In fig.\ref{rc2} we compare the influence of the cut
%                                   $\Delta > 30~GeV$
%w.r.t. to the case $E > 4~GeV$ (V.P.C. are included here).
%This cut reduces the size of the radiative corrections,
%  in particular                                        for the
%DA and the $e$ method which are brought
%                     to more controllable values.

%the DA method has relatively small r.c. when using a high $\Delta$
%cut, but divergent ones without it. The variation of the r.c.
%depends strongly on the $\Delta$ cut value.
%\item
%The $e$ method has r.c. which depends strongly
%on the value of Bjorken $x$ and on the $\Delta$ cut at low $x$.
%\item
%The $e\Sigma$ method has r.c. which are rather flat in $x$ at a given
%$Q^2$ and which are smaller than the $e$ or DA ones with the same cut
%conditions.
We can conclude that
the $\Sigma$ method has  small r.c.  and that they  barely
 depend on $x$, $Q^2$ or the presence of  the $\Delta$ cut,
%    contrarily to the $e$ and DA  methods,
    therefore allowing
%The $\Sigma$ method allows
                            to make a structure function
measurement  free of  radiative  corrections uncertainties.
           It is again complementary to the electron method which
has r.c. varying as a function of $x$, and also dependent on the
parton densities in the proton in the absence of a $\Delta$ cut.

\section{DA and $\Sigma$ Methods Independent of QED radiation}
\begin{figure}[htbp]
\epsfig{file=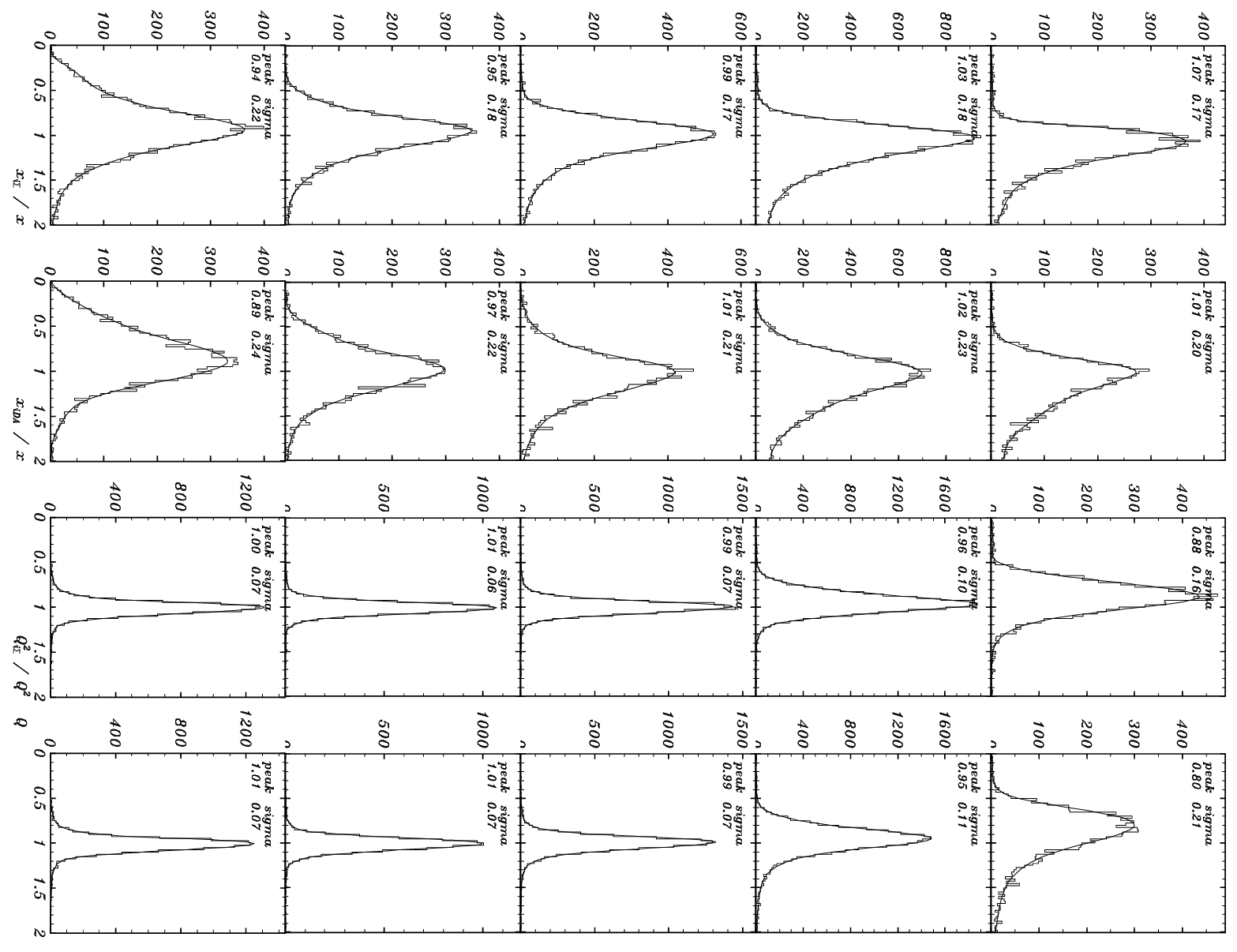,width=16cm,height=18cm,angle=90.}
\caption[]{\label{res5}
\sl Comparison of $x$ and $Q^2$ reconstruction at low $Q^2$
    ($Q^2 > 7~GeV^2$) for $\Sigma$ and DA method
    independent of initial QED-bremsstrahlung (I$\Sigma$,IDA)
                                 From top to bottom,
each row represent a bin in $y$:
                                                    very high
(0.5-0.8), high (0.2-0.5), medium (0.1-0.2), low (0.05-0.1), very low
(0.01-0.05).                                                     }
\end{figure}

In case of initial state radiation of a photon
from the electron we have
      $s_{true}=  4 (E^e-E^{\gamma})
E^p $. To fully constrain  the kinematics
we need  three independent variables and from the four variables
defined in section 2, four combinations are possible.
In practice only two
\footnote{The other two methods uses 2 variables from the hadronic
final state and only one from the electron, thus being less precise.}
                         will deserve here our attention: the
 two which come directly from the DA and the $\Sigma$ method.

The first of these methods was developped in
\cite{ben1}, by determining
%               modifying the double-angle
%method with the introduction of
         the initial energy $E^e$
from~$[E,\theta,\gamma]$.
%\begin{equation}
% E^e_{[\gamma E\theta]} =  E~\frac{\sin\theta + \sin\gamma -
%\sin(\theta+\gamma)}  {2\sin \gamma}
%\end{equation}
%A mixed method independent of initial state radiation
%was also studied by  introducing this definition of
%$E^e_{[\gamma E\theta]}$
%and showed satisfactorily results over the full $(x,Q^2)$
%plane~\cite{gbwh}.
%However this was not a satisfactory solution since
%it was based on the 4 variables ($E,\theta,\Sigma,\gamma$) and at this
%stage it would be better to use a kinematic fitting procedure.
Once $E^e$ is reconstructed, it can be injected in the DA method
to obtain   the formulae for the double-angle method
independent of ISR, labeled IDA here after:
\begin{equation}
E^e_{IDA}=
  E \ \frac{\tan\gamma /2+\tan\theta /2}{\cot\theta /2+\tan\theta /2}
                                               \hspace*{2cm}
x_{IDA} = \frac{E}{E^p}  \
      \frac{\cot\gamma /2+\cot\theta /2}{\cot\theta /2+\tan\theta /2}
\end{equation}
\begin{equation}
y_{IDA}=y_{DA} \hspace*{4cm}
Q^2_{IDA} = E^2 \ \tan \theta/2 \
    \frac{\tan\gamma /2+\tan\theta /2}{\cot\theta /2+\tan\theta /2}
\end{equation}

\noindent
The other method   is
 the $\Sigma$  independent
of ISR (I$\Sigma$), which is obtained by just using $s_{true}$
instead of $s$~\cite{ubgb}.
                We have $y_{I\Sigma} = y_{\Sigma}$, $Q^2_{I\Sigma}=
Q^2_{\Sigma}$ and
\begin{equation}
E^e_{I\Sigma} = \frac{ \Sigma + E (1-\cos\theta)} {2}
    \hspace*{2cm}
x_{I\Sigma} = \frac{ E}{E^p} \ \frac{\cos^2 \theta/2 }{  y_{\Sigma}}
\end{equation}
A comparison of the I$\Sigma$ and  IDA  methods is presented in
fig.6, following the scheme of figs.2-3. The following conclusions can
be drawn.
\begin{itemize}
\item
At low $Q^2$,
the I$\Sigma$ method is slightly more precise than the  IDA  one
in all $y$ bins both for  $x$ and for $Q^2$.
\item
$x_{I\Sigma}$ is less precise than
$x_{\Sigma}$ at high $y$, while they have similar resolutions
at low and medium $y$.
%hich is the price to pay when trying to reconstruct $E^e$.
\item
 $x_{IDA}$ and $Q^2_{IDA}$ are  more precise
than their DA counterpart  at low $x$, but have wider resolutions
at low $y$.
\end{itemize}
The loss of precision when trying to reconstruct $E^e$ is compensated
by the reconstruction of the true kinematics
when studying the special class of
           events with catastrophic initial state radiation.
However for the structure function measurement it is  better,
as it has been done on the 1992 and
                      1993 data by the H1 and ZEUS collaborations
\cite{H193,H194,ZEUS93,ZEUS94},                 to use
the methods described in the preceding sections ($e,m,DA,\Sigma$)
since the additional smearing observed in these two ISR independent
methods is larger than the smearing introduced by QED radiation.
These ISR independent methods can also be used as
has been demonstrated by a complete $F_2$ analysis  based
on the I$\Sigma$ method which has been done inside the H1
collaboration on the 1992 data~\cite{Wulf},
 in good agreement with the published measurements.
\section{Conclusion}
We presented a new prescription for DIS kinematic reconstruction
well adapted to study the low $x$-low $Q^2$ physics in the full $y$
range ($0.01 < y < 0.8$) and
        which allows to measure $F_2$ in the complete  kinematic domain
accessible at HERA. This $\Sigma$ method
can be modified to  become fully independent of ISR (I$\Sigma$ method)
although this is  not necessary, since the radiative
corrections for the $\Sigma$
        were shown to be already  very small, and some loss of
precision when using the  I$\Sigma$ method was observed.
By replacing $Q^2_{\Sigma}$ by $Q^2_e$ we defined
another  method which further
               optimizes the  reconstruction precision
 at the cost of becoming correlated to the $e$ method at high $y$.
For the
 kinematic fitting we recommended to consider the hadronic angle
instead of its transverse momentum as the second variable
(beside $\Sigma$) characterizing the hadronic final state.
After this  review on  the subject of
reconstruction of the deep inelastic scattering kinematics
                         at HERA, we  expect that
future improvements in this field will mostly come
from deeper knowledge and better use
                      of the detector response.

\vspace*{0.5cm}
\noindent
\begin{Large}
{\bf Acknowledgments}
\end{Large}

\vspace*{0.5cm}
\noindent
In this paper we relied extensively on the efforts of our H1
collaborators, with whom we built the software chain on which the
quantitative results presented here are based upon, and we would like
to thank them warmly. We also want to thank Johannes Bl\"umlein for his
interest in this work and for his great help on the radiative
correction part. We finally would like to thank Joel Feltesse
for usefull discussions and a careful reading of the manuscript.

\newpage
%noindent
\begin{Large}
{\bf Appendix:}
\end{Large}
   \hspace*{2.cm}   {\large $y,Q^2$  and $x$  formulae}

%\vspace*{0.5cm}
\begin{tabular}{|c|c|c|c|}
\hline
 &  &  &  \\
method &  $y$ & $Q^2$ & $x$ \\
 &  &  &  \\
\hline
 &  &  &  \\
 $ e $ &  $1-${\Large$\frac{E}{E^e}$}$\sin^{2}\frac{\theta}{2}$  &
          $ 4 E^eE  \cos^2\frac{\theta}{2}$  &
          $ Q^2/ys$  \\
 &  &  &  \\
 $ h $ &  {\Large $\frac{\Sigma}{2 E^e} $}&
%%%%%%%  {\Large $\frac{\Sigma^2}{\tan^2\frac{\gamma}{2} (1-y_h)}$} &
         {\Large $\frac{T^2     }{                        1-y_h }$} &
          $Q^2/ys $\\
 &  &  &  \\
 $ m      $ & $ y_h   $     &  $  Q^2_e $ &
          $Q^2/ys $\\
 &  &  &  \\
 DA &{\Large $\frac{\tan\gamma/2}{\tan\gamma/2+\tan\theta/2} $} &
     $4E^{e2}${\Large$\frac{\cot\theta/2}{\tan\gamma/2+\tan\theta/2}$}&
       $Q^2/ys $\\
 &  &  &  \\
 $\Sigma$&{\Large $\frac{\Sigma}{\Sigma+E(1-\cos\theta)}$}&
          {\Large $\frac{E^2\sin^2\theta}{1-y_{\Sigma}}$ }&
          $Q^2/ys $\\
 &  &  &  \\
  IDA    &   $ y_{DA}$ &
  $E^2 \ \tan\frac{\theta}{2}${\Large$\frac{\tan\gamma/2+\tan\theta/2}
           {\cot\theta/2+\tan\theta/2}$}  &
         {\Large $\frac{E}{E^p} \frac{\cot\gamma/2+\cot\theta/2}
          {\cot\theta/2+\tan\theta/2}$}\\
 &  &  &  \\
I$\Sigma$&$y_{\Sigma} $& $Q^2_{\Sigma} $&
   {\Large$\frac{ E}{E^p} \ \frac{\cos^2{\theta/2} }{ y_{\Sigma}} $}\\
 &  &  &  \\
\hline
\end{tabular}

%----------------------------------------------------------------------
%\newpage


\begin{thebibliography}{99}
%------- Introduction-------------------------
\bibitem{JoJo}
J.Feltesse,  ``Physics at HERA", vol. 1,
ed. R.D. Peccei, DESY (1987)~33-58

\bibitem{jb}
A.Blondel, F.Jacquet, Proceedings of the Study of an $ep$ Facility for
Europe, ed. U.Amaldi, DESY 79/48 (1979)~391-394

\bibitem{ben1}
S.Bentvelsen et al., Proceedings of the Workshop Physics at HERA,
vol. 1, eds. W. Buchm\"uller, G. Ingelman, DESY (1992)~23-40

\bibitem{hoe}
C. Hoeger, ibid., 43-55
%C.Hoeger, Proceedings of the Workshop Physics at HERA,
%vol. 1, eds. W. Buchm\"uller, G. Ingelman, DESY (1992)~43-55

\bibitem{gbwh}
G.Bernardi, W.Hildesheim, ibid., 79-100
%                          Proceedings of the Workshop Physics at HERA,
%vol. 1, eds. W. Buchm\"uller, G. Ingelman, DESY (1992)~79-100

\bibitem{maxu}
J. Bl\"umlein, M. Klein, Proceedings of the Snowmass Workshop ``The
Physics of the Next Decade'', ed. R. Craven, (1990)~549-551

\bibitem{H193}
H1 Collab., I. Abt et al., Nucl. Phys. B407 (1993)~515-535

\bibitem{bas}
U.Bassler, Ph.D. thesis, University of Paris VI, May 1993.


\bibitem{H194}
V. Brisson, LAL preprint 94-61 (1994),
to appear in the proc. of the 27th International Conference on High
Energy Physics, Glasgow (1994).

\bibitem{django}
G.A. Schuler, H. Spiesberger, Proceedings of the Workshop Physics at
HERA, vol. 3, eds. W. Buchm\"uller, G. Ingelman, DESY (1992)~1419-1432

\bibitem{heracles}
A. Kwiatkowski, H. Spiesberger and H.-J. M\"ohring, ibid., 1294-1310
%                                                   Physics at HERA,
%vol. 3, eds. W. Buchm\"uller, G. Ingelman, DESY (1992)~1294-1310

\bibitem{cdm}
L. L\"onnblad, Computer Phys. Comm. 71 (1992)~15-31
%\bibitem{cdmdis}
%  B. Andersson, G. Gustafson, L. L\"onnblad and U. Pettersson,
%  Z. Phys. C43 (1989) 625

\bibitem{h1flow}
H1 Collab., T. Ahmed et al., Phys. Lett. B298 (1993)~469-478

\bibitem{zeflow}
ZEUS Collab., M. Derrick et al., Z. Phys. C59 (1993)~231-242

\bibitem{mrsh}
%  A.D. Martin, W.J. Stirling and R.G. Roberts,
%  Phys. Lett. 306B (1993) 145, ibid. 309B (1993) 492.
A.D. Martin, W.J. Stirling, R.G. Roberts, Proceedings of the
Workshop on Quantum Field Theory Theoretical Aspects of High Energy
Physics, eds. B. Geyer and E.M.Ilgenfritz (1993)~11-26

\bibitem{geant}
R. Brun et al., GEANT3 User's Guide, CERN-DD/EE 84-1 (1987)

\bibitem{h1dete}
H1 Collab., I. Abt et al., DESY  93-103 (1993)
{\it submitted to Nucl. Instr. and Meth.}.

\bibitem{h1kine}
U.Bassler, G.Bernardi,  Kinematic Reconstruction inside H1,
H1-note/LPNHE-Paris preprint, {\it in litt.}



\bibitem{chaves}
H. Chaves, R.J. Seyfert, G. Zech, Proceedings of the Workshop Physics at
HERA, vol. 1, eds. W. Buchm\"uller, G. Ingelman, DESY (1992)~57-70

\bibitem{julian}
J.P. Phillips, Manchester preprint MAN/HEP/93/9 (1993), H1-note
 09/93-314 (1993)

\bibitem{spies}
H. Spiesberger et al., Proceedings of the Workshop Physics at HERA,
 vol.2,
eds. W. Buchm\"uller, G. Ingelman, DESY (1992)~ 798-838

\bibitem{blu}
J. Bl\"umlein, ibid., 1270-1284
%              Proceedings of the Workshop Physics at HERA, vol.3,
%eds. W. Buchm\"uller, G. Ingelman, DESY (1992)~ 1270-1284

J. Bl\"umlein, DESY 94-044 (1994), Z. Phys. C (1994) {\it in press}

\bibitem{uwe}
U. Obrock, private comunication.

\bibitem{ubgb}
U. Bassler, G. Bernardi, H1-note 09/93-274 (1993)

\bibitem{ZEUS93}
  ZEUS Collab., M. Derrick et al., Phys. Lett. B316 (1993)~412-426

\bibitem{ZEUS94}
ZEUS Collab., M. Derrick et al., DESY  94-117 (1994)
{\it submitted to Phys. Lett. B}.

\bibitem{Wulf}
N.Wulff, Ph.D. thesis, University of Hamburg, (1994)

\end{thebibliography}
\end{document}